\documentclass [10pt]{article}
\usepackage {amssymb}
\usepackage {amsmath}
\usepackage [cp1251]{inputenc}
\usepackage{graphicx}
\usepackage {longtable}

\begin{document}

\title{Energy Resolution Improvement of Scintielectron detectors:
Priorities and Prospects}

\author{ S.V. Naydenov \footnote{e-mail: naydenov@isc.kharkov.com}\: ,
V.D. Ryzhikov \footnote{e-mail: ryzhikov@stcri.kharkov.ua}
\\ {\it Institute for Single Crystals, National Academy of Sciences of
Ukraine } \\ {\it 60 Lenin Ave., 61001 Kharkov, Ukraine} }


\maketitle

\begin{abstract}
The development prospects of a scintillator-photodiode type
detector with an improved energy resolution attaining few per
cent, $R$ from $1$ to $2\% $, are considered. The main resolution
components have been analyzed theoretically, their theoretical and
physical limits have been established. Empirical data on the
properties of novel scintillators have been presented confir\-ming
the possibility of the $R$ improvement. New optimization methods
have been proposed to optimize the detector statistical
fluctuations and the scintillator intrinsic resolution. A specific
importance of the intrinsic resolution is shown as a limiting
threshold factor at the ionizing radiation energy values from
$662\, keV$ to $10\,MeV$ and over.
\end{abstract}

\textbf{I. INTRODUCTION}

Among the solid detectors, the scintillator-PMT (SC-PMT) and
scintillator- photodiode (SC-PD) type detectors are widely used
along with the semi\-conductor (SCD) ones. In the SCD, the
ionizing radiation is transformed directly into the charge
carriers. In contrast, in the SC-PMT and SC-PD ones, a two-stage
transformation takes place, first into optical photons (in the
scintillator) and then into the charge carriers (in the
photoreceiving device). The double transformation results in the
energy losses and redistribution (dispersion). Therefore, the
sensitivity and energy resolution of the SCD are one decimal order
higher. It is of a principal importance that, as the SCD volume
increases from $1\, mm^ 3$ to $1$ or $10\, cm^3$, these advantage
become lost due to the carrier re-capturing in the traps.
Therefore, the parameters of modern large-volume SCD are not so
high as those of scintielectron detectors at room temperatures.
The energy resolution of a SC-PD pair including the traditional
CsI(Tl) scintillator of a volume up to $1 cm^3$ and a silicon [1]
or HgI2 [2] photodiode at room temperature is about $5\% $ to $6\%
$ on the $662 keV$ line. The advance in the development of large
wide-band SCD is only insignificant during lasr few years. An
alternative way is to develop a novel high-efficiency scintillator
matched spectrally to a solid photoreceiver. This will provide a
detector with a volume of at least $10 cm^3$ having the energy
resolution of few per cent.

The theory of energy transformation process in scintillators is
developed well. There are several models to describe that process
and to calculate the energy resolution, e.g., [3--6] and others.
The possible correlation between important characteristics of the
process, such as the conversion efficiency, the light collection
coefficient, the self-absorption, etc., on the one hand, and the
energy resolution of the scintillation and of the entire system,
on the other one, is, however, still insufficiently studied. The
theoretical threshold of the energy resolution was not considered.
It is just the above-mentioned problems that are the consideration
objects in this work.

\textbf{II. ENERGY RESOLUTION COMPONENTS. \\
STATISTICAL CONTRIBUTIONS}

Let the energy resolution of a scintillator-photodiode couple be
considered. Under approximation of the Gaussian shape of the
electron signal output line, the resolution $R$ takes the form

\begin{equation}
\label{eq1} R = {\frac{{FWHM}}{{MAX}}} = G\,{\frac{{\Delta
E}}{{{\left\langle {E} \right\rangle} }}}\,;\quad G = \sqrt {8\ln
2} \approx 2{\kern 1pt} .{\kern 1pt} 355 \quad {\rm ,}
\end{equation}

\noindent where ${\left\langle {E} \right\rangle} $ is the
spectral line average energy; $\Delta E = { \left[ \left\langle
E^{2}\right\rangle - \left\langle E \right\rangle ^{2} \right]
}^{1/2}$, its dispersion. The depen\-dence on the radiation energy
$E_{\gamma} $ is defined by the system efficiency in the
photoabsorption peak, $\eta _{{\rm p}}$:

\begin{equation}
\label{eq2} E = \eta _{p} E_{\gamma}  ;\,{\left\langle {E}
\right\rangle}  = {\left\langle {\eta _{p}}  \right\rangle} {\kern
1pt} E_{\gamma}  ;\Delta E = f(E_{\gamma}  ;{\left\langle {\eta
_{p}}  \right\rangle} ;\Delta \eta _{p} ;...){\rm .}
\end{equation}

The conversion efficiencies of the system, namely, the total,
$\eta $ (the electron yield of the PD per $1 MeV$ of the ionizing
radiation), scintillation, $\eta _{{\rm s}{\rm c}}$ (the light
yield per $1 MeV$), and peak, $\eta _{{\rm p}}$ (the fraction of
the total efficiency $\eta $ falling on the photoabsorption peak),
ones, depend on the radiation energy as well as on a series of
physical and geometric parameters denoted in (\ref{eq2}) by dots
(...). These efficiencies are fluctuating quantities. That is why
there is no direct (monotonous) dependence of the line width,
$\Delta E$, and thus, of the detector resolution, $R=R(\eta )$, on
its conversion efficiency $\eta $ measured in the current regime.
However, the correlations associated with the light yield
improvement and its more efficient use are conserved. These
correlations result as a rule in a resolution improvement that is
confirmed by experimental data.

The fluctuation unmonotonity, $\Delta E=\Delta E(\eta )$, is due
to that the main parameters of the Gaussian distribution, namely,
the mean value and dispersion, are independent of each other.
Physically, this fact is due to that the energy transformation and
transfer processes are multifactorial and indefinite. The spectral
line widening does not result from the statistical contributions
only. The latter are related directly to the conversion efficiency
$\eta $. Non-statistical fluctuations are of no less importance.
The latter include the fluctuations of various geometric-dynamic
and spatially non-uniform factors influencing the detector
conversion output. Moreover, fluctuations due to that the light
yield is not in proportion with the energy $E_{\gamma} $ and those
of the mutually dependent (competing) quantities are associated
with the above-mentioned ones. Due to that complication, the most
efficient method to determine the resolution consists in that the
most important noise channels are separated is such a way that the
corresponding fluctuations may be assumed to be independent. Then,
each of those contributions is estimated. The independent
fluctuation are added vectorially [7]. Therefore, we have for the
total resolution (the index k is the noise channel number):

\begin{equation}
\label{eq3} R^{2} = {\sum\limits_{k} {R_{k}^{2}} } {\rm .}
\end{equation}

It is naturally that the resolution of a scintielectron detector
can be determined as

\begin{equation}
\label{eq4} R^{2} = R_{sc}^{2} + R_{st}^{2} + R_{pd}^{2} {\rm ,}
\end{equation}

\noindent where $R_{sc}$ is the scintillator intrinsic resolution;
$R_{st}$, the statistical fluctuations of the energy carriers
(photons and electrons); $R_{pd}$, the noise of the photodiode and
electronic devices. Note that each contribution in the formula
(\ref{eq4}) may contain in its turn partial components
characterizing the specific widening mechanisms. Since $R_{k} \le
R$ for each partial channel , it is necessary to attain the noise
level of at least $1$ to $2\% $ for each component to provide the
same value of the total resolution.

Let the conditions be determined allowing us to neglect the
photodiode noise. This contribution depends mainly on the total
number of noising electrons $N_{noise}$ relative to the useful
electronic signal

\begin{equation}
\label{eq5} R_{pd} \propto R_{noise} \propto {\frac{{N_{noise}}
}{{N_{e}} }} \approx {\frac{{(\Delta E)_{pd}} }{{\eta E_{\gamma} }
}} \quad {\rm ,}
\end{equation}

\noindent where $\Delta E_{pd}$ is the corresponding energy
spread; $\eta $ means the mean value of the conversion efficiency,
if not otherwise specified. It will be shown below that a
sufficiently high $\eta $ (of the order of $10\% $) is required to
compensate the statistical detector noise. Then at the mean energy
values of $0.5$ to $1 MeV$, the useful signal $N_{e}$ should be at
least $30 000$ electrons. It follows from the expression
(\ref{eq5}) that the photodoide resolution will be smaller than
$1--2\% $ on condition that the spread $(\Delta E)_{pd}$ does not
exceed $1--2 keV$ or $N_{noise}$ is less than $200$ or $300$
``dark'' electrons. Such a low noise level has been attained
already in modern semiconductor photodiodes, see [8] and other
publications. For example, the specific silicon photodiode being
used as the SCD exhibits the resolution of $5.5 keV$ on the $122
keV$ line of $^{{\rm 5}{\rm 7}}$Co and detects reliably the $5.5
keV$ signal from $^{{\rm 5}{\rm 5}}$Fe [8]. Thus, its resolution
on the $^{{\rm 1}{\rm 3}{\rm 7}}$Cs line expressed in relative
units should amount $2.36\times (5.5 keV/662 keV)  \propto 0.8\%$,
that is, less than $1\% $. The photodiode noise is even lower in
the case of high energy spectrometry, since $R_{pd} \propto
N_{e}^{ - 1} \propto E_{\gamma} ^{ - 1} $ in the Eq. (\ref{eq5}).
Thus, the electronic noises may be assumed to be suppressed
strongly, at least under forced cooling. Their level $R_{pd} $ is
lower than $1\% $. Then it is just the statistical noises that are
of decisive importance in the moderate energy spectroscopy.

The statistical resolution $R_{st}$ consists of independent
contributions of quantum fluctuations of the number (or energy) of
the scintillation photons $R_{st,ph}$ and the photodiode
photoelectrons $R_{st,el}$:

\begin{equation}
\label{eq6} R_{st}^{2} = R_{st,ph}^{2} + R_{st,el}^{2}
\end{equation}

In contrast to the cascade theory [9] for the SC-PMT assembly, the
statistical fluctuations of photons cannot be neglected in this
case. For SC-PD with a small $R_{st}$, a nearly ideal energy
transport from the scintillation quanta to the photodiode photons
is required. Therefore, the statistical noises of the photons and
electrons will be of the same order of magnitude. The $R_{st,ph}$
contribution is often neglected also in SC-PD detectors, but at
worse resolution values exceeding $4--5\% $.

Let a formula be derived for $R_{st}$ in a detector with a high
conversion efficiency. The $\gamma $-quantum energy transformation
into a scintillation photon and then into an electronic signal
will be considered to that end as independent events within a
Bernoulli's test series [7]. The statistical fluctuations of such
a test series are described by a binomial distribution with the
mean value $x = N\,p$ and the dispersion $Dx = N\,p(1 - p)$. Here
$N$ it the number of tests, i.e., the number of converted
particles at $100\% $ efficiency. The event probability, $p$,
means the scintillation efficiency $\eta _{sc}$ and the total one
$\eta $, respectively. Thus, the partial statistical contributions
will be expressed as

\begin{equation}
\label{eq7} R_{st,{\kern 1pt} j} = 2.\;355\,\;\sqrt {{\frac{{1 -
\eta _{j}} }{{\eta _{j} N_{j}} }}} \,;\;N_{j} = {\frac{{E_{\gamma}
} }{{\varepsilon _{j}} }};\;j = (sc;\,el){\rm ,}
\end{equation}

\noindent where $\varepsilon _{sc}$ and $\varepsilon _{el}$ are
the mean energy of the scintillation quantum and the mean energy
of an electron-hole pair formation in the semiconductor,
respectively. The total statistical contribution is

\begin{equation}
\label{eq8} R_{st} = {\frac{{2.{\kern 1pt} 355}}{{\sqrt {\eta
\,\left( {E_{\gamma}  / \varepsilon _{el}}  \right)}} }}\;{\left[
{1 + K_{Y} \,\left( {1 - \eta _{sc} \,{\frac{{\varepsilon _{sc} +
\varepsilon _{el}} }{{\varepsilon _{sc} }}}} \right)} \right]}^{1
/ 2}{\rm ,}
\end{equation}

\noindent where the coefficient $K_{Y} = \left( {\eta / \eta
_{sc}}  \right)\,\left( {\varepsilon _{sc} / \varepsilon _{el}}
\right)$ is introduced. Usually, $\varepsilon _{sc} = hc / \lambda
= 1239 / \lambda (nm) \propto 2\div 3 eV$ and, for example,
$\varepsilon _{el}(Si)=3.6 eV$ in a silicon SCD. In detectors with
low conversion yields, in the limiting case $\eta \le \eta _{sc} <
1$, the usual Poisson noise of the charge carriers is obtained:

\begin{equation}
\label{eq9} R_{st} \approx R_{st,el}^{poisson} =
{\frac{{2.355}}{{\sqrt {\eta \,\left( {E_{\gamma}  / \varepsilon
_{el}}  \right)}} }} \quad {\rm .}
\end{equation}

At $\eta _{sc} $ exceeding $20\% $ or $\eta $ not less than $10\%
$, the distinctions of the binomial distribution from the Poisson
one become substantial. The fluctuations of the former drop
steeper as those of the latter. This points directly that an
increase in $\eta _{sc} $ is of prospects for the maximum
attenuation of the threshold statistical fluctuations. Under ideal
energy transformation when $\eta = \eta _{sc} = 1$, the
theoretical limit is attained:

\begin{equation}
\label{eq10} R_{st}^{ideal} = R_{st} \left( {\eta = 1} \right) = 0
\quad {\rm .}
\end{equation}

An important feature of the statistical resolution follows from
the expression (\ref{eq8}). This resolution is characterized by a
monotonous dependence on both conversion efficiencies. The other
contributions to the total resolution do not exhibit that
property. It is useful to transform the Eq. (\ref{eq8}) into the
form

\begin{equation}
\label{eq11} R_{st} = {\frac{{2.{\kern 1pt} 355}}{{\sqrt {\eta
_{sc} \;\left( {E_{\gamma } / \varepsilon _{el}}  \right)}}
}}\;\sqrt {K_{Y}^{ - 1} + \left( {1 - \eta _{sc}
\,{\frac{{\varepsilon _{sc} + \varepsilon _{el}} }{{\varepsilon
_{sc} }}}} \right)} {\rm .}
\end{equation}

Solving this equation for $\eta _{sc} $, we obtain

\begin{equation}
\label{eq12} \eta _{sc} = \left( {{\frac{{6.\,91}}{{\lambda
\,E_{\gamma} } }}} \right)\,\;{\frac{{1 + K_{Y}^{ - 1}} }{{{\left[
{R_{st}^{2} + \left( {{\frac{{6.\,91}}{{\lambda \,E_{\gamma} } }}}
\right)\;{\frac{{\varepsilon _{sc} + \varepsilon _{el}}
}{{\varepsilon _{el}} }}} \right]}}}} \quad {\rm ,}
\end{equation}

\noindent where the energy $E_{\gamma}$ should be expressed in
($keV$) and the scintillation wavelength $\lambda $, in ($nm$).
This expression defines the scintillation efficiency necessary to
provide the preset statistical noise level in a detector with the
coefficient $K_{Y}$ defining the matching between the scintillator
and the photodiode, $0 < K_{Y} < 1$. At the ideal spectral and
optical matching, the $K_{Y}$ value is defined mainly by the light
collection coefficient $K_{c}$, so that $(\varepsilon _{el} /
\varepsilon _{sc} )K_{Y} \approx K_{c} $. For a ``red''
scintillator with $\lambda =640 nm$, at a good light collection
with $K_{c}= 0.6$ to $0.8$, the statistical resolution at the $662
keV$ energy is small on the condition following from the Eq.
(\ref{eq12}),

\begin{equation}
\label{eq13} R_{st} (E_{\gamma}  \propto 1\,MeV) \le 1\% \;
\Leftrightarrow \;\eta _{sc} \ge \,\;25\% \quad {\rm .}
\end{equation}

For the total conversion efficiency \textit{$\eta $}, that level
will amount $10$ to $15\% $. In the energy region up to $10 MeV$
and over, the statistical resolution drops dramatically, since
$R_{st} \propto 1 / \sqrt {E_{\gamma} }  $. In the best
scintillation assemblies, the statistical noise at the $662 keV$
line is from $2\% $ to $4\% $, see., e.g., [10] and other
publications. At high $E_{\gamma} $ from $10 MeV$ to $1 GeV$, this
contribution will decrease by a factor from $3-4$ to several
decimal orders and amounts several fractions of a percent.

Thus, the statistical contribution can be minimized substantially
in a detector with a high scintillation efficiency and
well-matched SC-PD couple. The development of new scintillators
based on semiconductors (e.g., ZnSe(Te) [10]) or rare-earth
elements (e.g., LaCl3(Ce) [11]) evidences the real attaining of
extremely small $R_{st} $values even in the moderate energy
region. The high light yield is necessary also to optimize the
intrinsic energy resolution of the scintillator, $R_{sc} $. It
includes usually non-statistical fluctuations. This quantity is a
natural threshold of the limiting improved detector resolution.
This is associated with that $R_{sc}$ includes residual (slightly
dependent on or independent at all of $E_{\gamma}$) contributions
at any radiation energy, including the high one. Its part will be
decisive in the improved-spectrometry detectors.

\textbf{III. THE SCINTILLATOR INTRINSIC RESOLUTION}

The decisive role of the intrinsic resolution noted in [3--6],
[8--11] and other works is clearly pronounced in alkali halide
scintillators where it amounts from $4\% $ to $5\% $. In heavy
oxides BGO and CWO the internal resolution of sufficiently small
samples (less than $10 cm^3$ in volume) is negligible, since their
light yield is in proportion to the ionizing radiation energy.
Note that recent studies, in particular, in [12--15], confirm
convincingly the dominating dependence between the light yield
non-proportionality and the deterioration of the detector
intrinsic resolution. The non-proportionality is due mainly to the
conversion efficiency non-linearity with respect to the secondary
electron formation at the radiation absorption by the scintillator
and to uncorrelated processes of the multiple Compton's
scattering. The proportionality effect is manifested in numerous
experimental data. For example, RbGd2Br7(Ce) seems to exhibit a
rather good intrinsic resolution at the proportional light yield
[13]; its total resolution at the $662 keV$ line is $4.1\% $, the
statistical contribution of the PMT statistical noise of $3.5\% $
being the main component in this case. The intrinsic resolution of
ZnSe(Te) scintillator amounts $3.26\% $ at the same line of
$^137$Cs while the total resolution is $5.37\% $ [14].

The intrinsic resolution includes several components. Some of
those drop monotonously depen\-ding on $E_{\gamma}$. There are
components, however, independent of or slightly depending on the
ionizing radiation energy provided that it is absorbed completely
in the crystal (the escape resolution and edge effects being
neglected). Among those threshold contributions, the substantional
resolution of the scintillator $R_{sub}$ and the light collection
resolution $R_{lc}$ are most substantial, so that

\begin{equation}
\label{eq14} R_{sc}^{2} = R_{sub}^{2} + R_{lc}^{2} + R_{r}^{2}
\quad {\rm ,}
\end{equation}

\noindent where the insubstantial rest of the intrinsic resolution
is denoted as $R_{r}$. To neglect the escape resolution, it is
necessary to use the crystals having the characteristic dimension
$L$ not less than

\begin{equation}
\label{eq15} L \ge L_{e} \approx {\frac{{0.{\kern 1pt} 45{\kern
1pt} E_{\gamma} (MeV)}}{{\rho {\kern 1pt} \,(g / cm^{3})}}} \quad
{\rm ,}
\end{equation}

\noindent where $L_{e} $ is the radiative $\delta $-electron free
path in the radiation energy region under consideration. The edge
effects can be neglected in ``heavy'' crystals of a volume of
several tens of $mm^3$ or more. To retain a small crystal volume
at very high energy levels, metallized reflectors are to be used.
Other advantages offered by the mirror scintillator boundary will
be considered below. The substantional resolution $R_{sub}$ is due
mainly to the light yield non-proportionality, $E_{sc} =\eta _{sc}
E_{\gamma}$; $\eta _{sc} = \eta _{sc}(E_{\gamma})\ne const$, and
by its spatial inhomogeneity. The light collection contribution
$R_{lc}$ is defined by the geometric-dynamic fluctuations in the
crystal of a preset shape at fixed optical parameters in the
crystal volume and at its boundary.

The spatial inhomogeneity of scintillations, $\eta _{sc}=\eta
_{sc} \left( {\vec {r}} \right)$, is of great importance, in
particular, in activated compounds. The resolution of
inhomo\-geneities is defined by the factor $R_{inhom} =
2.36(\Delta \eta _{sc})/{\left\langle {\eta _{sc}} \right\rangle}$
with the spatial averaging of the corresponding fluctuations. To
suppress the statistical noise, a high mean scintillation
efficiency ${\left\langle {\eta _{sc}} \right\rangle} $ of about
$25\% $ is necessary at moderate energies. To attain the low
macroinhomo\-geneity resolution in this case, the dispersion
$\Delta \eta _{sc}$ should not excess $0.1\% $. It is just a
superhigh homogeneity of the activator distribution that answers
to this requirement.

There are scintillators with extra small $R_{sub} $ values. Their
specific feature is the light yield proportionality. Those include
the above- mentioned tungstates. For example, CdWO4 of $200\,cm^3$
volume has $R_{sub} $ less than $0.3\%$ (after this contribution
is isolated from the total resolution) and $0.03\%$ to $0.08 \%$
at the crystal volume from $3\,cm^3$ to $20\,cm^3$ (data of [16]
were used for the estimations). ZnSe(Te) shows a rather good
linearity with $\eta _{sc}(5.9 keV)/\eta _{sc}(662keV) =85\% $ and
$\eta _{sc}(16.6keV)/\eta _{sc}(662keV)=90\% $ [14], the physical
light yield being $28 000 ph/MeV$. Some complex oxides behave
somewhat worse. So, Lu3Al5O12(Ce) has the light yield $13 000\,
ph/MeV$ and $\eta _{sc}(16.6keV)/\eta _{sc}(662keV)=76\% $, while
LuAlO3(Ce) where the light yield is decreased down to $11 000\,
ph/MeV$ shows $\eta _{sc}(16.6 keV)/\eta _{sc}(662 keV)=71\% $
[15]. To compare, the non-linearity of NaCl(Tl) amounts about
$80\%$ at the light yield of $40 000\, ph/MeV$. Nevertheless, the
above-mentioned modern scintillators and other ones offer good
prospects in the attaining of the high scintillator energy
resolution (both intrinsic and total one) as their spectrometric
characteristics will be further improved.

To minimize the substantional contribution, it is necessary to
develop a material with a high scintillation efficiency as well as
high light yield proportionality and homogeneity. The
scintillators exhibiting the intrinsic (or nearly intrinsic)
luminescence type seem to be of priority. There is no strict
theory explaining in terms of physical phenomena why unactivated
scintillators or those activated with isovalent impurities show
substantially improved substantional resolutions. This regularity,
however, is observed in experiments. This may be due to the
following physical reasons. The absence of a non-isovalent
activator, on the one hand, provides the uniform and homogeneous
distribution of the emission centers over the sites of the ideal
crystal lattice. On the other hand, the absence of the direct
energy transformation where its intermediate transport from the
matrix to the emission centers is required is accompanied by
losses and results in the detected pulse widening.

The scintillators containing non-isovalent impurities (emission
centers) show a considerable non-proportionality of the light
yield, in particular, near the $K$-absorption threshold. For
example, among yttrium garnets, it is just crystals containing the
isovalent cerium admixture, YAlO3(Ce) (YAP) and Y3Al5O12(Ce) (YAG)
that exhibit the best proportionality while Y2SiO5(Ce) (YSO) with
non-isovalent Si and Ce the worse one, cf. [13]. The
proportionality of Lu2SiO5(Ce) (LSO) and Gd2SiO5(Ce) (GSO) is
rather poor. In the same time, it is considerably better at zinc
selenide with isovalent tellurium or at lanthanum chloride with
isovalent cerium. Furthermore, tungstates and germanates with
their intrinsic emission centers in the crystal matrix show an
ideal proportionality. Non-activated NaI and CsI also exhibit a
smaller intrinsic resolution than those activated with sodium or
thallium. Moreover, at low temperatures (liquid nitrogen) pure NaI
and CsI have considerably higher light yields relative to NaI(Tl)
amounting $217\% $ and $190\% $, respectively [16]. The light
yield is increased when the interaction with the lattice phonons
is suppressed. The considerable increase of the scintillation
efficiency can be associated with the more pronounced definiteness
of the energy transformation in the absence of activator. This
should be accompanied by an improvement in the substantional
resolution. It is just what is observed even at room temperatures
when the light yield of activated compounds is enhanced. This is
due to the same cause, a very small intrinsic resolution of
non-activated scintillators. Finally, it is to note that it is
just the PbWO4 (PWO) scintillator that has been chosen as the main
component of high-energy calorimeters for the CERN accelerator and
other ones. This is connected, not in the last place, with its
extremely low substantional resolution $R_{sub} $. At high
energies, $1 GeV$ to $100 GeV$ and over, a resolution smaller than
1\% has been already attained for such detectors.

The geometric fluctuations of the light collection, $R_{lc}$,
under homogeneous scintillation distri\-bution are defined only by
the light collection coefficient dispersion. This resolution
component, in spite of its non-statistical character, has the
following form in the Gaussian approximation:

\begin{equation}
\label{eq16} R_{lc} = 2.{\kern 1pt} 355\,{\frac{{\sqrt
{{\left\langle {K_{c}^{2}} \right\rangle}  - {\left\langle {K_{c}}
\right\rangle} ^{2}} }}{{{\left\langle {K_{c}}  \right\rangle} }}}
\quad {\rm .}
\end{equation}

The light collection resolution and the coefficient $K_{c} $
itself depend on the scintillator geometric shape and optical
properties, namely, the light reflection, refraction and
absorption. The resolution attains its minimum under ideal light
collection with mirror reflection at the boundary and without
light absorption in the scintillator.

The reason for the ideality of the mirror light collection is
established in the frame of the stochastic (geometric-dynamic)
light collection theory [17]. The picture of the geometric (light)
rays distribution in a detector of macroscale size can be
substituted by a dynamic model, a billiard with elastic
reflections from the boundary. The billiards are described by
special dynamic systems, the reversible maps in a symmetric phase
space with coordinates $\left( {\varphi _{1} ,\varphi _{2}}
\right)$. A couple of such coordinates defines a ray with two
successive reflection points at the billiard (detector) boundary.
The light collection parameters are expressed in terms of the
invariant distribution function$ f$ for the corresponding dynamic
flow. The latter defines the total set of all possible light rays
being reflected from the billiard boundary. It is just the
possibility to consider the light collection picture not for
individual trajectories but in its entirety that makes a
substantial distinction of this approach from widespread numerical
models. The non-averaged quantities $K_{c} $ and $K_{c}^{2} $ are
expressed as

\begin{equation}
\label{eq17} K_{c} = \int\!\!\!\int {f\left( {\varphi _{1}
,\varphi _{2}}  \right)\chi _{c} \left( {\varphi _{1} ,\varphi
_{2}}  \right)} A\left( {\varphi _{1} ,\varphi _{2}}
\right)\,d\varphi _{1} \,d\varphi _{2} {\rm ;}
\end{equation}

\begin{equation}
\label{eq18} K_{c}^{2} = \int\!\!\!\int {f\left( {\varphi _{1}
,\varphi _{2}} \right)\chi _{c}^{2} \left( {\varphi _{1} ,\varphi
_{2}}  \right)} A^{2}\left( {\varphi _{1} ,\varphi _{2}}
\right)\,d\varphi _{1} \,d\varphi _{2} {\rm ,}
\end{equation}

\noindent where $\chi _{M} $ is the characteristic function of the
set $M$ (equal to unity in the points of the set and zero outside
of it); $\chi _{c} $ corresponds to the light pick-off in the
phase space that does not include the captured light region. The
factor $A \propto \exp \,\left( { - \mu _{sc} \,L} \right)$
describes the light losses; $\mu _{sc} $ is the optical absorption
coefficient of the scintillator; $L$ -- the length of the latter.
The expressions (\ref{eq17}) and (\ref{eq18}) should be averaged
over the flash distribution in the scintillator, depending on the
corresponding distribution (that was assumed above to be
homogeneous). For the mirror light collection with mirror
boundaries, $r \approx 1$; neglecting the absorption in an
optically transparent crystal, we have $A \approx 1$. Using the
projection property $\chi ^{2} = \chi $, we obtain the equality
${\left\langle {K_{c}^{2}} \right\rangle}  = {\left\langle {K_{c}}
\right\rangle} $. Then, starting from (\ref{eq16})--(\ref{eq18}),
we obtain the relationship

\begin{equation}
\label{eq19} R_{lc}^{mirror} = 2.\,355\,\sqrt {{\frac{{1 -
{\left\langle {K_{c}} \right\rangle} }}{{{\left\langle {K_{c}}
\right\rangle} }}}} \quad {\rm .}
\end{equation}

Mathematically, the obtained fluctuation character corresponds to
the binomial distribution. At the ideal light collection, when
$K_{c} = 1$ and $r = 1$, the theoretical limit is attained:

\begin{equation}
\label{eq20} R_{lc} (K_{c} = 1{\kern 1pt} ;\;r = 1) = 0 \quad ,
\end{equation}

\noindent that is confirmed by numerical and experimental data
(see e.g. [10], [16]). The limiting high $K_{c} $ values are
attainable in detectors where the dynamics of light rays is
completely chaotic. Due to the ``disjoining of correlations'', the
captured light is absent therein and the technical light yield is
limited by the absorption factor only. The mentioned detector
shape answers to the chaotic billiard geometry. The known
geometries of those type include the ``stadium'', the cube with an
internal void, etc. As an exception, it is just sphere
corresponding to the integrable billiard that is characterized by
the ideal light collection.

The light collection becomes changed considerably if the
absorption takes place ($A \ne 1$). This results not only in a
reduced intensity of the scintillation propagation but causes the
light collection inhomogeneity. The effective light collection
with a high resolution is attained in small-size scintillators
and/or in those having a high optical transparency. It follows
from experimental data, numerical calculations and theoretical
estimations that a scintillator of regular geometry with a volume
of about $10 cm^3$ (e.g., a cylinder of commen\-surable height and
diameter values about $3 cm$) and a high-quality mirror boundary
($r$ from $0.8$ to $0.9$) will exhibit a sufficiently low light
collection resolution on condition that

\begin{equation}
\label{eq21} R_{lc} \le \;1\% \; \Leftrightarrow \;\mu _{sc}
\,L\,\; \le \,\;0.\,1\quad \left( {10L\,\; \le \,\;l_{sc}}
\right) \quad {\rm ,}
\end{equation}

\noindent where $l_{sc} = \mu _{sc}^{ - 1} $ is the light ray free
path in the scintillator. On the same conditions, a good light
collection $K_{c} $, up to $60 \% $ , is attained. It is just the
alkali halide crystals that exhibit a high transparency, $l_{sc}
$[NaI(Tl)] attaining $2 m$; it is somewhat lower in tungstates,
$l_{sc}$ being from $10 cm$ to $50 cm$. Nevertheless, the $l_{sc}$
values of about $30 cm$ that are required for a scintillator of
about $10 cm^3$ volume are quite attainable. Note that the optical
transparency increase improves, in an indirect manner, the
substantional resolution $R_{sub} $ that depends also on the
energy loss in the scintillator.

The increase of the scintillator volume results as a rule in a
worse intrinsic energy resolution due to the optical absorption.
But exceptions are possible when a special geometry is chosen that
optimizes the light collection. The search for scintillators with
high atomic numbers, $Z_{eff} $, and density values, $\rho $,
becomes very actual. This approach allows to avoid the sharp
increase of the scintillator volume in the moderate and high
energy spectrometry. The improvement of the light collection and
its resolution is a geometric-dynamic problems to a considerable
extent and does not require any substantial changes in the
scintillator production technology. Basing on the modern advances
in the non-linear physics, some untraditional scintillator
geometries could be developed, such as asymmetrical shapes,
varying curvature boundaries, ``stadium'' type defocusing
billiards, systems containing topological defects, e.g., holes.
The light collection inhomogeneities must be optimized in such
scintillators due to strong stochastic mixing of the light rays.

Thus, the light collection intrinsic resolution is minimized in
mirror-reflecting scintillators of appropriate geometry and high
optical transparency. Alternatively, volume-diffuse systems
consisting of small crystalline grains in an optically conductive
medium can be used for scintillators with high optical losses. The
intrinsic material resolution is expected to be small in
non-activated scintillators or ``combined'' ones (where the
emission of an isovalent activator and the intrinsic one of the
scintillator are combined) at a high conversion efficiency,
homogeneity and the complete absorption of the ionizing radiation.
Scintillators with a high $Z_{eff} $and meeting the condition
$\left( {2\div 3} \right)\,l_{r} \le L$, where $l_{r} $ is the
radiation attenuation length. The intrinsic scintillator
resolution can be lowered down to a level of $1-2\% $ by
satisfying the above-mentioned conditions in combination.

\textbf{IV. CONCLUSIONS}

It follows from the above theoretical analysis that there are no
physical reasons hindering to attain sufficiently small values of
the main energy resolution components in a SC-PD detector. Only in
the low energy range (several hundreds $keV$) where it is just the
statistical contribution that predominates, the latter increases
sharply as the radiation energy drops even in scintillators with a
limiting scintillation quantum yield. In contrast, the $R$ values
from $15$ to $2\% $ are quite attainable in the range of $662 keV$
to $10 MeV $ and over. To that end, it is necessary to increase
substantially the total conversion efficiency ($1.5$ to $2$ times)
and the scintillation efficiency (up to $25$ to $30\% $) as
compared to the available SC-PD detectors having $R$ of $3$ to
$5\% $. Moreover, optically transparent, homogeneous scintillators
exhibiting a high light yield proportionality, in particular those
close to the proper emission type, are to be searched for. To
improve the light collection, the search for untraditional
scintillator geometry (e.g., the ``stadium'' type one) providing a
strong chaotization of light beams therein (at the mirror surface)
should be preferred. The last advances in the field of ``heavy''
crystals with light yields exceeding that of CsI(Tl) (see
[10],[11], [18-20], etc.) evidence clearly that these problems are
quite realizable and offer good prospects.

\textbf{V. REFERENCES}

[1] J. M. Markakis, ``Energy resolution of CsI(Tl) crystals with
various photoreceivers'', IEEE Trans. Nucl. Sci., NS-35, no. 1, p.
356, 1988.

[2] J.M. Markakis, ``High resolution scintillation spectroscopy
with HgI$_{{\rm 2}}$ as photodetector'', Nucl. Instrum. and Meth.,
vol. A 263, no. 2-3, p. 499, 1988.

[3] P. Iredale, ``The effect of the nonproportional responce of
NaI(Tl) crystals to electrons upon to resolution of gamma rays'',
Nucl. Instrum. and Meth., vol. 11, p. 340, 1961.

[4] R.B. Murray and A. Meyer, ``Scintillation response of
activated inorganic crystals to various charged particles'',
Physical Review, vol. 122, no. 3, p. 815, 1961.

[5] C.D. Zerby, A. Meyer and R.B. Murray, ``Intrinsic line
broadening in NaI(Tl) gamma-ray spectrometers'', Nucl. Instrum.
and Meth., vol. 12, p. 115, 1961.

[6] R. Hill, A.J.L. Collinson, ``The variation of resolution with
gamma energy in activated alkali halide crystals'', Proc. Prys.
Soc., vol. 85, p. 1067, 1965.

[7] W. Feller, ``An Introduction to Probability Theory and Its
Applications'', vol. 1, New York, 1970.

[8] H. Grossman, Nucl. Instrum. and Meth., vol. A 295, no. 3, p.
400, 1990; A. Gramsch, K. Lynn, M. Weber et al., Nucl. Instrum.
and Meth., vol. A311, no. 3, p. 529, 1992.

[9] E. Bretenberger, Progr. Nucl. Phys., v.10, no. 4, p. 16, 1955.

[10] L.V Atroshchenko, S.F. Burachas, L.P Gal'chinetskii, B.V.
Grinyov, V.D. Ryzhikov, N.G. Starzhynski, ``Scintillator crystals
and detectors of ionizing radiations on their base'', Kyiv,
Naukova Dumka, 1998 (in Russian).

[11] E.V.D. van Loef, P. Dorenbos, C.W.E. van Eijk et al.,
``High-energy-resolution scintillator: Ce$^{{\rm 3}{\rm +} }$
activated LaCl3'', Applied Phys. Letters, vol. 77, no. 10, p.
1467, 2000.

[12] J.D. Valentine, B.D. Rooney, J. Li, ``The light yield
nonproportionality component of scintillator energy resolution'',
IEEE Trans. Nucl. Sci., vol. 45, p. 512, 1998.

[13] P. Dorenbos, J.T.M. de Hass, C.W.E. van Eijk,
``Nonproportionality in the scintillation response and the energy
resolution obtainable with scintillation crystals'', IEEE Trans.
Nucl. Sci., vol. 42, no. 1, p. 2190, 1995; P. Dorenbos, M.
Marsman, C.W.E. van Eijk, ``Energy resolution non-proportionality,
and absolute light yield of scintillation crystals'', Inorganic
Scintillators and Their Applications, Proc. of the Intern. Conf.
SCINT95, ed. by P. Dorenbos, C.W.E. van Eijk, Delft, the
Netherlands, 1995, p. 148; J.C. Guilliot-Noel, J.C. van't Spijker,
J.T.M. de Hass, P. Dorenbos, C.W.E. van Eijk, ``Scintillation
properties of RbGd$_{{\rm 2}}$Br$_{{\rm 7}}$(Ce). Advantages and
limitations'', IEEE Trans. Nucl. Sci., vol. 46, no.5, p. 1274,
1999.

[14] M. Balcerzyk, M. Moszynski and M. Kapusta, ``Energy
resolution of contemporary scintilla\-tors. Quest for high
resolution, proportional detector'', Proceedings of the 5$^{{\rm
t}{\rm h}}$ Intern. Conf. on Inorganic Scintillators and Their
Applications, SCINT99, Moscow, August 16-20, 1999, p. 167-172.

[15] M. Balcerzyk, M. Moszynski, M. Kapusta et al., ``YSO, LSO,
GSO and LGSO. A study of energy resolution and
non-proportionality'', IEEE Trans. Nucl. Sci., vol. 47, no. 4, p.
1319, 2000.

[16] Yu. A. Tsirlin, M.E. Globus, E.P. Sysoeva, ``Optimization of
detecting gamma-irradiation by scintillation crystals'',
Energoatomizdat, Moscow, 1991 (in Russian).

[17] S.V. Naydenov, V.V. Yanovsky, ``Stochastical theory of light
collection. I. Detectors and billiards'', Functional Materials,
vol. 7, no.4 (\ref{eq2}), p. 743-752, 2000; ``II. Projective
geometry and invariant distribution function of a billiard'',
Ibidem, vol. 8, no.1, p. 27-35, 2001.

[18] S. Derenzo, W. Moses, ``Experimental efforts and results in
finding new heavy scintillators'', Heavy scintillators. Proc. of
the Intern. Conference ``Crystal2000'', (ed. by F. Notarisstefani
et al.), Chamonix, France, 1992, p. 125.

[19] W.W. Moses, M.J. Weber, S.E. Derenzo et al., ``Recent results
in a search for inorganic scintillators for X- and gamma ray
detection'' Proc. of the Intern. Conference on Inorganic Scint.
and Their Applications, SCINT97, (ed. by Y. Zhiwen et al.),
Shanghai, China, September 22-25, 1997, p. 358; J.C. van't
Spijker, P. Dorenbos, C.P. Allier, C.W. E.van't Eijk, A.E. Ettema,
``Lu$_{{\rm 2}}$S$_{{\rm 3}}$(Ce$^{{\rm +} {\rm 3}}$): a new red
luminescing scintillator'',. Ibidem, p. 311; J.C. van't Spijker,
P. Dorenbos, C.W.E.van't Eijk et al., ``Scintillation properties
of LiLuSiO4(Ce$^{{\rm + }{\rm 3}}$)'', Ibidem, p. 326.

[20] E.V.D. van Loef, P.Dorenbos, and C.W.E. van Eijk,
``High-energy-resolution scintillator: Ce$^{{\rm 3}{\rm +} }$
activated LaBr$_{{\rm 3}}$'', Appl. Phys. Lett., vol. 79, no. 10,
p. 1573, 2001.

\end{document}